\documentstyle[twoside,fleqn,espcrc2]{article}

\newcommand{\be}{\begin{eqnarray}}
\newcommand{\ee}{\end{eqnarray}}
\newcommand{\ben}{\begin{eqnarray*}}
\newcommand{\een}{\end{eqnarray*}}

\newcommand{\lla}{\left\langle}
\newcommand{\rra}{\right\rangle}

\def\diffn#1	  {\Delta^{-}_{#1}}
\def\mb#1       {\mbox{\boldmath $#1$}}

\def\preprints{
\vspace{-14ex}
{\small
\begin{tabbing}
\` {\sl hep-lat/9808051} \\
    \\
\` LSUHE No. 271--1998 \\
\` August 1998 \\
\end{tabbing} 
}
\vspace*{0.1in}
}

\title{
\preprints
Dynamically generated electric charge distributions in Abelian 
projected  SU(2) lattice gauge theories.}
\author{Alistair Hart, Richard W. Haymaker\thanks{Presented 
        by R. Haymaker.
   Work partially supported by the U. S. Department of Energy.}
   and Yuji Sasai\thanks{Supported in part by the Japan
   Ministry of Education. 
   permanent address: Oshima National College of Maritime 
   Technology, 1091-1, Oshima-cho, Oshima-gun, Yamaguchi 743-21, Japan}.
   \\
   \vspace{0.5ex}
   \noindent Dept. of Physics and Astronomy, 
       	Louisiana State University, Baton Rouge, Louisiana,
       	70803  USA        }
       	
\begin{document}

\begin{abstract}
We show in the maximal Abelian gauge  the dynamical 
electric charge density generated by the coset fields, gauge
fixing and ghosts  shows antiscreening as in the case of 
the non-Abelian charge.   We verify that
with the  completion of the ghost term all contributions to
flux are accounted for in an exact lattice Ehrenfest relation. 
\end{abstract}

\maketitle


Lattice studies based on Abelian projection have had 
considerable success identifying
the dynamical variables relevant to the physics of
quark confinement. There is no definitive way as yet
of choosing the optimum variables, but in the 
maximal Abelian gauge 
\cite{thooft81,kronfeld87a}
the U(1) fields remaining after Abelian projection
produce a heavy quark potential that continues to rise
linearly 
\cite{suzuki90}.
Further the string tension is almost, but not exactly,
equal to the full SU(2) quantity; 92\% in a recent study
at $\beta = 2.5115$
\cite{bali96}.

All elements of a dual superconducting vacuum appear to
be present 
\cite{mandelstam76,thooft81};
in the maximal Abelian gauge  
magnetic monopoles reproduce nearly all of the U(1) string
tension
\cite{stack94,bali96}.
The spontaneous breaking to the U(1) gauge symmetry 
is signalled by the
non-zero vacuum expectation value of monopole operator
\cite{chernodub96,digiacomo98}.
The profile of the electric field and the 
persistent magnetic monopole currents 
in the vortex between quark and antiquark  
are well described by an
effective theory, the Ginzburg--Landau, or equivalently a Higgs
theory giving a London penetration depth and Ginzburg--Landau 
coherence length
\cite{singh93,bali98}.

Central to finding the effective theory is the definition of
the field strength operator in the Abelian projected theory,
entering not only in the vortex profiles but also in the
formula for the monopole operator. All definitions should be
equivalent in the continuum limit, 
but use of the appropriate lattice
expression should lead to a minimization of discretization
errors.

In Ref.
\cite{dhh98} 
we exploit lattice symmetries to derive such an
operator that satisfies Ehrenfest relations; Maxwell's 
equations for ensemble averages irrespective of
lattice artifacts. 

The charged coset fields are normally discarded in Abelian
projection, as are the ghost fields arising from the 
gauge fixing procedure. Since the remainder of the SU(2)
infrared physics must arise from these, an understanding of
their r{\^o}le is central to completing the picture of full
SU(2) confinement.  In the maximal Abelian gauge a 
localised cloud of like polarity charge is
induced in the vacuum in the vicinity of a source, producing
an effect reminiscent of the antiscreening of charge in 
{\sc QCD}. In other gauges studied, the analogous current
acts to {\em screen} the source
\cite{bdh}. 
(This is a tentative result, however, without the benefit of 
the refined definition of flux.)

%
%
%
%
%

Consider the effect of a `right shift' 
of a particular link, 
$U_\mu(x_0) \to U_\mu(x_0) U^s(x_0)$:
\ben
Z_W(\{U^s\}) = \int [d (UU^s)]\; W_3(U)\;\Delta_{FP} \; 
\delta[F]\; e^{\beta S(U)},
\een
where we have introduced
\ben
1 = \Delta_{FP}  \int \prod_{j,y} d g_j(y)  \prod_{i,x}
 \delta[F^g_i(U^{\{g_j(y)\}}; x)],
\een
and integrated out the $g$ variables in the standard way. 
So $\Delta_{FP} = \det M$ where
\ben
M_{ix;jy} = \left. \frac{\partial F^g_i(x)}{\partial g_j(y)}
\right|_{g=0}
\een
We take as
the source term an SU(2) plaquette with a $\sigma_3$ insertion to 
check the theorem.
\ben
W_3 \equiv \frac{1}{2}Tr (U^{\dagger}U^{\dagger} U U i \sigma_3).
\een
$Z_W$  is {\em not invariant} under $U^s$.  The shift 
is inconsistent with the gauge condition.   
It is invariant, however, under an infinitessimal shift together 
with an infinitessimal `corrective' gauge transformation
that restores the gauge fixing
\ben
U^s(x_0) = 1 - \frac{i}{2}\epsilon_{3}(x_0) \sigma_3; \;
G(x) = 1 - \frac{i}{2}\mb{\eta} (x) \cdot  \mb{\sigma} . 
\een
Using the invariance of the measure under combination of
a shift and a `corrective' gauge transformation we obtain
\ben
\left[ \frac{\partial}{\partial \epsilon_{\mu}(z_0)} + \sum_{k,z} \;\;
\frac{\partial \eta_k(z)}{\partial \epsilon_{\mu}(z_0)}
\frac{\partial }{\partial \eta_k(z)}
\right]Z_W = 0.
\een
\begin{table}[tb]
\begin{tabular}{lll} 
\hline \hline
Source:\hspace{1cm} 
& 
$W_3$  \hspace{1cm}
& 
$W_3(U \rightarrow D)$ 
\\ 
Ehrenfest term 
&
&
\\ 
\hline
$\lla (W_3)_{\mu}\biggr|_s \rra$
&
0.65468(10) \hspace{1cm}&  0.63069(20) 
\\ \\
$\lla (W_3)_{\mu}\biggr|_g \rra$
&  
0.06095(7) \hspace{1cm}&  0.04463(4) 
\\ \\
$\lla W_3  \frac{(\Delta_{FP})_{\mu}}{\Delta_{FP}}\biggr|_g \rra$
&  
0.00127(21) \hspace{1cm}&  0.00132(50) 
\\ \\
$\lla W_3  \frac{(\Delta_{FP})_{\mu}}{\Delta_{FP}}\biggr|_s \rra$
&  
0.00529(3) \hspace{1cm}&  0.00564(3)  
\\ \\
$\lla \beta (S)_{\mu}\biggr|_s \rra$
&  
-0.72246(68) \hspace{1cm}&  -0.68275(50)
\\ 
\hline
Zero  & -0.00026(77)& -0.00045(64) \\
\hline \hline
\\ 
\end{tabular}
 
\caption{Terms in the Ehrenfest relation, Eqn.(1) on a
$4^4$ lattice at $\beta = 2.5$.  The column
labeled $W_3$ corresponds to the source described in the text.
In the second column the source links are replaced by their
diagonal parts
of the links to test a second source. The theorem gives
zero for the sum.}

\label{table_one}

\vspace{-2ex}
\end{table}

\noindent
The Ehrenfest relation reads (where $( \cdots)_{\mu}$ indicates a 
derivative
\cite{dicecio97})
\be
&&\lla (W_3)_{\mu}\biggr|_s + 
(W_3)_{\mu}\biggr|_g +   W_3 \times \right. \nonumber \\  
   && \left.  \left(
\frac{(\Delta_{FP})_{\mu}}{\Delta_{FP}}\biggr|_s + 
\frac{(\Delta_{FP})_{\mu}}{\Delta_{FP}}\biggr|_g  +
\beta (S)_{\mu}\right)
\rra  = 0.
\ee
Assuming the source involved the shifted link,
\begin{itemize}

\item 
$(W_3)_{\mu}\biggr|_s =   \pm
\frac{1}{2}Tr (U^{\dagger}U^{\dagger} U U) $.

\item 
$(W_3)_{\mu}\biggr|_g$ comes from the corrective gauge 
transformation acting on the source which is U(1) invariant but
not SU(2) invariant. 

\item 
$\frac{(\Delta_{FP})_{\mu}}{\Delta_{FP}}\biggr|_s$ is the effect of
the shift on the Faddeev-Popov determinant.

\item 
$\frac{(\Delta_{FP})_{\mu}}{\Delta_{FP}}\biggr|_g$ is due to
the corrective gauge transformation of the Faddeev-Popov determinant.

\item 
$\beta (S)_{\mu}$ is a shift term of the (gauge invariant) action.

\end{itemize}
Imposing the gauge constraint
up to first order quantifies $\eta$
\ben
F_i(x) + \frac{\partial F_i(x)}{\partial \epsilon_\mu(z_0)} 
\epsilon_\mu(z_0)
+ \sum_{k,z} \frac{\partial F_i(x)}{\partial \eta_k(z)} 
\eta_k(z) \equiv 0,
\een
and we define the shifted Faddeev-Popov matrix as a derivative
with respect to a general gauge transformation of the corrected 
constraint.  
\ben
&&M_{i x; j y} + \delta M_{i x; j y}  = 
\frac{\partial}{\partial g_j(y)}  \times 
\\ 
&&\left\{
F^g_i(x) + \frac{\partial F^g_i(x)}{\partial \epsilon_\mu(z_0)} 
\epsilon_\mu(z_0)
+ \sum_{k,z} \frac{\partial F^g_i(x)}{\partial \eta_k(z)} 
\eta_k(z)
\right\}. 
\een
Finally we evaluate the derivative using
\ben
\frac{(\Delta)_{\mu}}{\Delta}  = Tr[ M^{-1} (M)_{\mu}].
\een
\begin{table}[bt]
\begin{tabular}{cccl} 
\hline \hline
$\beta$ 
&
$\frac{1}{\beta}$
&
$div E$ 
& 
total flux  
\\ 
& 
& 
(on source)
\\ 
\hline
10.0 &  0.1 & 0.1042(1) &  0.0910(8) (mid) \\ 
&&& 0.0148(8) (back)\\ 
&&& 0.1092(8) (total)\\ 
\hline
2.4 &  0.4166 & 0.5385(19) &  0.7455(70) (mid) \\ 
&&& 0.0359(72) (back)\\ 
&&& 0.7815(95) (total) \\
\hline \hline
\\
\end{tabular}

\caption{$div E \equiv \langle \diffn{\nu} F_{\nu 4} \rangle$, 
normalized to $\frac{1}{\beta}$ for a `classical' point charge,
measured on a $3 \times 3$ Wilson loop source on an $8^4$ lattice.  
Integrated electric flux is measured on 
the midplane centered
on the Wilson loop and on a plane on
the far side of the torus, and the sum being the total flux.}

\label{table_two}

\vspace{-2ex}
\end{table}

\noindent
A check of this Ehrenfest theorem is given in Table~\ref{table_one}. 
Some of the terms require a 
$ 2 N \times 2 N $ matrix inversion,  where $N$ is the lattice 
volume. Hence we chose a $4^4$ lattice for the numerical test of
what is an exact relation on all lattice sizes.

We separate the links $U_{\mu}$ into diagonal 
$D_{\mu}$ and off-diagonal $O_{\mu}$
parts. Grouping all $O_{\mu}$ terms on the right as a set of 
conserved currents
we get the final form of the Ehrenfest-Maxwell  relation:
\ben
&& \lla \Delta_{\mu} F_{\mu \nu} \rra = \lla J_{\nu}^{dyn.} \rra
+ J_{\nu}^{static}\biggr|_{s} \nonumber \\ &&
+ J_{\nu}^{static}\biggr|_{g}
+ \lla J_{\nu}^{FP}\biggr|_{s}\rra
+ \lla J_{\nu}^{FP}\biggr|_{g}\rra.
 \een
The first term in the current comes from the excitation of the
charged coset fields,  the static term has an extra non-local
contribution coming from the corrective gauge transformation, 
and the last two contributions are from the ghost fields. 
These terms give a non vanishing charge density cloud around
a static source.  The left hand side can be used as a lattice
operator to measure the total charge density and
does not require the matrix inversions needed to measure the 
individual terms separately
which limited the numerical tests to small lattices.

Table~\ref{table_two} gives an application showing: \newline 
(i) a `classical' point charge
is dressed with like charge,  \newline
(ii) the total integrated flux is 
larger than $div E$ on the source, both indicating anti-screening.

In summary the coset fields renormalise 
the charge of the Wilson loop as measured by 
$\langle \diffn{\nu} F_{\nu 4} \rangle$
and charge is also induced in the surrounding vacuum. Full
SU(2) has antiscreening/asymptotic freedom of color
charge, and in the maximal Abelian gauge alone we have  seen
analogous behaviour, in that the source charge is increased 
and induces charge of like polarity in the neighboring vacuum.
The improved field strength expression
defined by the Ehrenfest identity does not coincide with the lattice
version
\cite{bdh} 
of the 't Hooft field strength operator
\cite{thooft74}.

\end{document}